\documentclass[pre,twocolumn,aps]{revtex4}
\newcommand{\bec}[1]{\mbox{\boldmath $ #1$}}
\usepackage{graphicx}
\begin{document}
\bigskip
\bigskip
\title{\bf Nonlinear shear-current dynamo and magnetic
helicity transport in sheared turbulence}
\author{IGOR ROGACHEVSKII}
\email{gary@bgu.ac.il}
\author{NATHAN KLEEORIN}
\author{EDIK LIVERTS}
\affiliation{\rm Department of Mechanical Engineering, The
Ben-Gurion University of the Negev, POB 653, Beer-Sheva 84105,
Israel}
\date{Received  7 July 2006; in final form 25 August 2006}

\begin{abstract}
The nonlinear mean-field dynamo due to a shear-current effect in a
nonhelical homogeneous turbulence with a mean velocity shear is
discussed. The transport of magnetic helicity as a dynamical
nonlinearity is taken into account. The shear-current effect is
associated with the ${\bf W} {\bf \times} {\bf J}$ term in the mean
electromotive force, where ${\bf W}$ is the mean vorticity due to
the large-scale shear motions and ${\bf J}$ is the mean electric
current. This effect causes the generation of large-scale magnetic
field in a turbulence with large hydrodynamic and magnetic Reynolds
numbers. The dynamo action due to the shear-current effect depends
on the spatial scaling of the correlation time $\tau(k)$ of the
background turbulence, where $k$ is the wave number. For Kolmogorov
scaling, $\tau(k) \propto k^{-2/3}$, the dynamo instability occurs,
while when $\tau(k) \propto k^{-2}$ (small hydrodynamic and magnetic
Reynolds numbers) there is no the dynamo action in a sheared
nonhelical turbulence. The magnetic helicity flux strongly affects
the magnetic field dynamics in the nonlinear stage of the dynamo
action. Numerical solutions of the nonlinear mean-field dynamo
equations which take into account the shear-current effect, show
that if the magnetic helicity flux is not small, the saturated level
of the mean magnetic field is of the order of the equipartition
field determined by the turbulent kinetic energy. Turbulence with a
large-scale velocity shear is a universal feature in astrophysics,
and the obtained results can be important for elucidation of origin
of the large-scale magnetic fields in astrophysical sheared
turbulence.

\bigskip


\end{abstract}

\maketitle

\section*{1. Introduction}

It is generally believed that one of the main reasons for generation
of large-scale magnetic fields in turbulent flow is the $\alpha$
effect (see, e.g., books and reviews by Moffatt 1978, Parker 1979,
Krause and R\"{a}dler 1980, Zeldovich {\it et al.} 1983, Ruzmaikin
{\it et al.} 1988, Stix 1989, Roberts and Soward 1992, Ossendrijver
2003, Brandenburg and Subramanian 2005a). However, the $\alpha$
effect caused by the helical random motions of conducting fluid,
requires rotating inhomogeneous or density stratified turbulence.

In a turbulence with a large-scale velocity shear and high
hydrodynamic and magnetic Reynolds numbers there is a possibility
for a mean-field dynamo (see Rogachevskii and Kleeorin 2003, 2004).
Turbulence with a large-scale velocity shear is a universal feature
in astrophysical plasmas. The large-scale velocity shear creates
anisotropic turbulence with a nonzero background mean vorticity
${\bf W}$. This can cause the ${\bf W} {\bf \times} {\bf J}$ effect
(or the shear-current effect), which creates the mean electric
current along the original mean magnetic field and produces the
large-scale dynamo even in a nonrotating and nonhelical homogeneous
turbulence (see Rogachevskii and Kleeorin 2003, 2004). Here ${\bf
J}$ is the mean electric current.

The mean-field dynamo instability is saturated by the nonlinear
effects. A dynamical nonlinearity in the mean-field dynamo which
determines the evolution of small-scale magnetic helicity, is of a
great importance due to the conservation law for the total (large
and small scales) magnetic helicity in turbulence with very large
magnetic Reynolds numbers (see, e.g., Kleeorin and Ruzmaikin 1982,
Gruzinov and Diamond 1994, 1996, Kleeorin {\it et al.} 1995, 2000,
2002, 2003a, 2003b, Kleeorin and Rogachevskii 1999, Blackman and
Field 2000, Vishniac and Cho 2001, Blackman and Brandenburg 2002,
Brandenburg and Subramanian 2005a, Zhang {\it et al.} 2006). On the
other hand, the effect of the mean magnetic field on the motion of
fluid and on the cross-helicity can cause quenching of the mean
electromotive force which determines an algebraic nonlinearity. The
combined effect of the dynamic and algebraic nonlinearities
saturates the growth of the mean magnetic field.

\begin{figure}
\vspace*{2mm} \centering
\includegraphics[width=7cm]{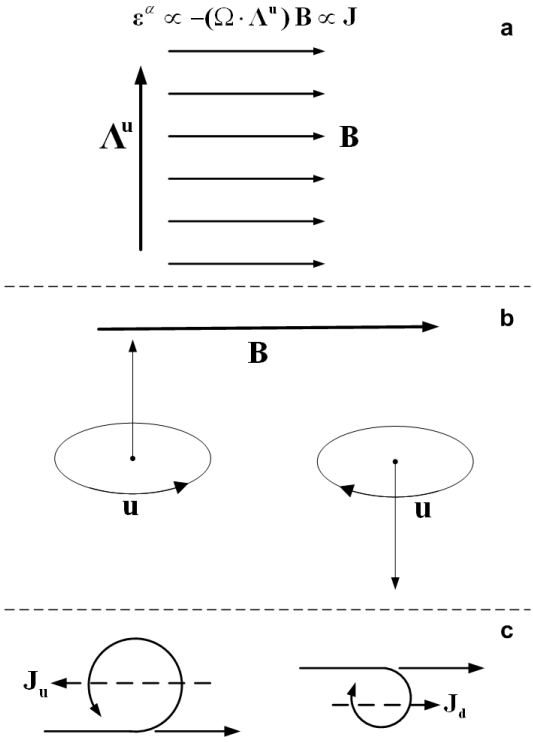}
\caption{\label{Fig1} Mechanism for the $\alpha$ effect: (a).
Interaction between uniform original mean magnetic field and
inhomogeneous rotating turbulence; (b). The deformations of the
original magnetic field lines are caused by the upward and downward
turbulent eddies; (c). Formation of the mean electric current
$\bec{\cal E}^\alpha \equiv \alpha {\bf B} \propto - l_0^2 \, ({\bf
\Omega} {\bf \cdot} {\bf \Lambda}^u) {\bf B} \propto {\bf J}$
opposite to the original mean magnetic field (for ${\bf \Omega} {\bf
\cdot} {\bf \Lambda}^u > 0)$. Here ${\bf J}_u$ and ${\bf J}_d$ are
the electric currents caused by the deformations of the original
magnetic field lines by the upward and downward turbulent eddies,
respectively, and $|{\bf J}_u| > |{\bf J}_d|$.}
\end{figure}

\begin{figure}
\vspace*{3mm} \centering
\includegraphics[width=7cm]{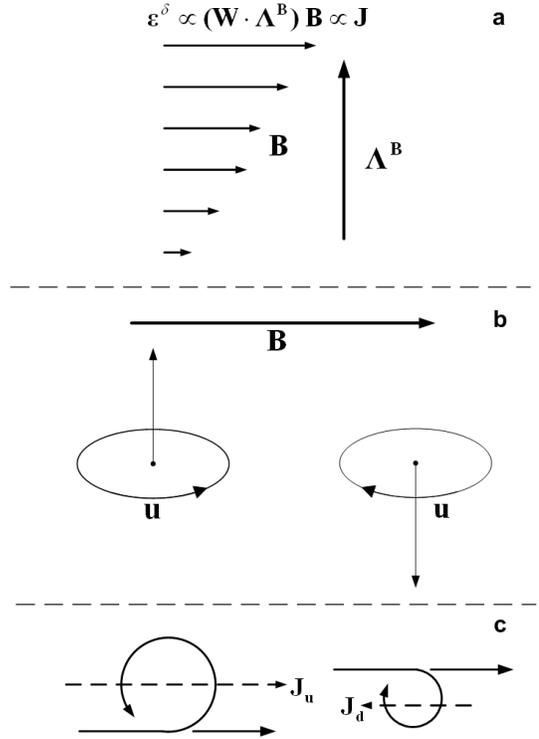}
\caption{\label{Fig2} Mechanism for the ${\bf W} {\bf \times} {\bf
J}$ effect: (a). Interaction between nonuniform original mean
magnetic field and homogeneous sheared turbulence; (b). The
deformations of the original magnetic field lines are caused by the
upward and downward turbulent eddies; (c). Formation of the mean
electric current $\bec{\cal E}^\delta \propto - l_0^2 \, {\bf W}
{\bf \times} (\bec{\nabla} {\bf \times} {\bf B}) \propto l_0^2 \,
({\bf W} {\bf \cdot} {\bf \Lambda}^B) {\bf B} \propto {\bf J}$ along
the original mean magnetic field (for ${\bf W} {\bf \cdot} {\bf
\Lambda}^B>0)$. Here ${\bf J}_u$ and ${\bf J}_d$ are the electric
currents caused by the deformations of the original magnetic field
lines by the upward and downward turbulent eddies, respectively, and
$|{\bf J}_u| > |{\bf J}_d|$.}
\end{figure}

The mean-field dynamo is essentially nonlinear due to the evolution
of the small-scale magnetic helicity (Gruzinov and Diamond 1994,
1996). In particular, even for very small mean magnetic field the
magnetic $\alpha$ effect is not small. This is a reason why we have
to take into account the dynamical nonlinearity in the mean-field
dynamo. When we ignore the dynamical nonlinearity due to evolution
of small-scale magnetic helicity and take into account only
algebraic nonlinearity caused by the nonlinear shear-current effect,
we obtain the saturated level of mean magnetic field which is in
several times larger than the equipartition field determined by the
turbulent kinetic energy (Rogachevskii {\it et al.} 2006). This
result can be important in view of astrophysical applications
whereby the super-equipartition large-scale magnetic fields are
observed, e.g., in the outer parts of a few galaxies (see Beck 2004,
2005). Note that it is a problem to reach a super-equipartition
level of the large-scale magnetic field in the $\alpha\Omega$
dynamo.

 The goal of this study is to investigate the nonlinear mean-field
dynamo due to the shear current effect. In this study we have taken
into account the dynamic nonlinearity caused by the evolution of the
small-scale magnetic helicity. This paper is organized as follows.
In section~2 we elucidate the physics of the shear-current effect.
In section~3 we consider kinematic dynamo problem due to this
effect. In section~4 we describe the results of numerical solutions
of the nonlinear mean-field dynamo equations which take into account
the shear-current effect, the algebraic and dynamic nonlinearities.
Finally, the discussion and conclusions are given in section~5. In
Appendixes~A we compare the problem of generation of the mean
magnetic field in a turbulence with a large-scale velocity shear
with that of the generation of mean vorticity in a sheared
turbulence.

\section*{2. The physics of the shear-current effect}

In this section we elucidate the mechanism of generation of
large-scale magnetic field due to the shear-current effect. To this
end we first discuss the physics of the $\alpha$ effect (see, e.g.,
Moffatt 1978, Parker 1979, Krause and R\"{a}dler 1980, Zeldovich et
al. 1983, Ruzmaikin {\it et al.} 1988). The $\alpha$ term in the
mean electromotive force $\bec{\cal E} = \langle {\bf u} \times {\bf
b} \rangle$ in a rotating inhomogeneous turbulence can be written in
the form $ \bec{\cal E}^\alpha \equiv \alpha {\bf B} \propto - l_0^2
\, ({\bf \Omega} {\bf \cdot} {\bf \Lambda}^u) {\bf B}$ (see, e.g.,
Krause and R\"{a}dler 1980, R\"{a}dler {\it et al.} 2003), where
${\bf u}$ and ${\bf b}$ are fluctuations of the velocity and
magnetic field, respectively, angular brackets denote ensemble
averaging, ${\bf \Omega}$ is the angular velocity, the vector $ {\bf
\Lambda}^u = \bec{\nabla} \langle {\bf u}^2 \rangle / \langle {\bf
u}^2 \rangle $ determines the inhomogeneity of the turbulence, ${\bf
B}$ is the mean magnetic field and $l_0$ is the maximum scale of
turbulent motions (the integral turbulent scale). The $\alpha$
effect is caused by the kinetic helicity $\chi_u \propto \eta_{_{T}}
\, ({\bf \Omega} {\bf \cdot} {\bf \Lambda}^u) $ in an inhomogeneous
rotating turbulence, where $\eta_{_{T}} \propto l_0 \, u_0$ is the
turbulent magnetic diffusion and $u_0$ is the characteristic
turbulent velocity in the maximum scale of turbulent motions $l_0$.
The deformations of the magnetic field lines are caused by upward
and downward rotating turbulent eddies (see figure~1). The
inhomogeneity of the turbulence breaks a symmetry between the upward
and downward eddies. Therefore, the total effect of these eddies on
the mean magnetic field does not vanish, and it creates the mean
electric current along the original mean magnetic field due to the
$\alpha$ effect.

The large-scale magnetic field can be generated even in a
nonrotating and nonhelical turbulence with a mean velocity shear due
to the shear-current effect (see Rogachevskii and Kleeorin 2003,
2004). This effect is related to the ${\bf W} {\bf \times} {\bf J}$
term in the mean electromotive force, and it can be written in the
form $\bec{\cal E}^\delta \propto - l_0^2 \, {\bf W} {\bf \times}
(\bec{\nabla} {\bf \times} {\bf B}) \propto l_0^2 \, ({\bf W} {\bf
\cdot} {\bf \Lambda}^B) {\bf B}$, where the mean vorticity ${\bf W}
= \bec{\nabla} {\bf \times} {\bf U}$ is caused by the mean velocity
shear and $ {\bf \Lambda}^B = \bec{\nabla} {\bf B}^2 / 2{\bf B}^2 $
determines the inhomogeneity of the mean original magnetic field. In
a sheared turbulence the inhomogeneity of the original mean magnetic
field breaks a symmetry between the influence of upward and downward
turbulent eddies on the mean magnetic field. The deformations of the
original magnetic field lines in the ${\bf W} {\bf \times} {\bf J}$
effect are caused by the upward and downward turbulent eddies. This
creates the mean electric current along the mean magnetic field and
produces the magnetic dynamo (see figure~2).

\begin{figure}
\vspace*{2mm} \centering
\includegraphics[width=7cm]{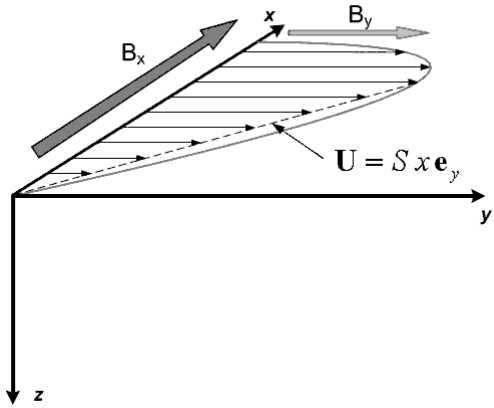}
\caption{\label{Fig3} Mechanism for shear-induced generation of
perturbations of the mean magnetic field $B_y$ by sheared stretching
of the field $B_x$. This effect is determined by the first term $
(\propto S B_x) $ in RHS of Eq.~(\ref{E3}), and it is similar to the
differential rotation because $ \bec{\nabla} {\bf \times} ({\bf U}
{\bf \times} {\bf B}) = S B_x {\bf e}_y $.}
\end{figure}

In order to demonstrate how the shear-current dynamo operates, let
us consider a homogeneous turbulence with a mean velocity shear,
${\bf U} = (0, Sx, 0)$ and $ {\bf W} = (0,0,S)$. Let us assume that
the mean magnetic field has a simple form ${\bf B} = (B_x(z),
B_y(z), 0)$. The mean magnetic field in the kinematic approximation
is determined by
\begin{eqnarray}
{\partial B_x \over \partial t} &=& - \sigma_{_{B}} \, S \, l_0^2 \,
B''_y + \eta_{_{T}} \, B''_x  \;,
\label{E2}\\
{\partial B_y \over \partial t} &=& S \, B_x + \eta_{_{T}} \, B''_y
\;, \label{E3}
\end{eqnarray}
(Rogachevskii and Kleeorin 2003, 2004), where $B''_i =
\partial^2 B_i / \partial z^2 $, $\eta_{_{T}}$ is the coefficient of
turbulent magnetic diffusion and the dimensionless parameter
$\sigma_{_{B}}$ determines the ${\bf W} {\bf \times} {\bf J}$ effect
(see Eq.~(\ref{M2}) below). The first term $ \propto S B_x $ in the
right hand side of Eq.~(\ref{E3}) determines the stretching of the
magnetic field $B_x$ by the shear motions, which produces the field
$B_y$ (see figure~3). On the other hand, the interaction of the
non-uniform magnetic field $B_y$ with the background vorticity ${\bf
W}$ (caused by the large-scale shear) produces the electric current
along the field $B_y$. This implies generation the field component
$B_x$ (see figure~4) due to the ${\bf W} {\bf \times} {\bf J}$
effect, which is determined by the first term in the right hand side
of Eq.~(\ref{E2}). This causes the dynamo instability.

\begin{figure}
\vspace*{2mm} \centering
\includegraphics[width=7cm]{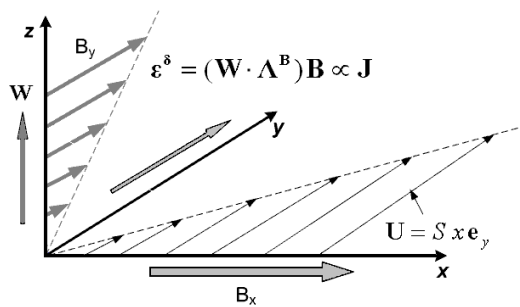}
\caption{\label{Fig4} Mechanism for shear-current generation of
perturbations of the mean magnetic field $B_x$ from the
inhomogeneous magnetic field $B_y$. This effect is determined by the
first term in RHS of Eq.~(\ref{E2}).}
\end{figure}

\section*{3. Kinematic dynamo due to the shear-current effect}

Let us consider the kinematic dynamo due to the shear-current
effect. In this study we have derived a more general form of the
parameter $\sigma_{_{B}}$ for arbitrary scaling of the correlation
time $\tau(k)$ of turbulent velocity field. The generalized form of
the parameter $\sigma_{_{B}}$ defining the shear-current effect, is
derived using Eq.~(A44) given in the paper by Rogachevskii and
Kleeorin (2004). The parameter $\sigma_{_{B}}$ entering in
Eq.~(\ref{E2}) is given by
\begin{eqnarray}
\sigma_{_{B}} = {4 I_0 \over 15} \, \biggl[1 + {I \over I_0} + 3 \,
\epsilon \biggr] \;, \label{M2}
\end{eqnarray}
where
\begin{eqnarray}
I = \int \tau'(k) \, k \, \tau(k)\, E(k) \, dk  \;, \quad I_0 = \int
\tau^2(k) \, E(k) \, dk \;,
\nonumber\\
\label{VE4}
\end{eqnarray}
$E(k)$ is the turbulent kinetic energy spectrum, $\tau(k)$ is the
scale-dependent correlation time of turbulent velocity field, the
parameter $\epsilon = E_m \, l_m / E_v \, l_0$, $\, E_m$ and $E_v$
are the magnetic and kinetic energies per unit mass in the
background turbulence (with a zero mean magnetic field), $l_m$ is
the scale of localization of the magnetic fluctuations generated by
the small-scale dynamo in the background turbulence and $\tau'(k)
=d\tau / dk$. Equations~(\ref{M2}) and~(\ref{VE4}) are written in
dimensionless form, where the turbulent time $\tau(k)$ is measured
in the units of $\tau_0= l_0/u_0$, the wave number $k$ is measured
in the units of $l_0^{-1}$, and the turbulent kinetic energy
spectrum $E(k)$ is measured in the units of $u_0^{2} \, l_0$.

The solution of equations~(\ref{E2}) and~(\ref{E3}) we seek for in
the form $ \propto \exp(\gamma \, t + i K_z \, z) ,$ where the
growth rate, $\gamma$, of the mean magnetic field due to the
magnetic dynamo instability is given by
\begin{eqnarray}
\gamma = S \, l_0 \, \sqrt{\sigma_{_{B}}} \, K_z - \eta_{_{T}} \,
K_z^2 \; . \label{IM10}
\end{eqnarray}
The necessary condition for the magnetic dynamo instability is
$\sigma_{_{B}} > 0$. The dynamo action due to the shear-current
effect depends strongly on the spatial scaling of the correlation
time $\tau(k)$ of the turbulent velocity field. In particular, when
$\tau(k) \propto k^{-\mu}$, the ratio $I/I_0=-\mu$, and the
criterion for the dynamo instability reads
\begin{eqnarray}
1 - \mu + 3 \, \epsilon > 0 \; . \label{M10}
\end{eqnarray}
For example, when $\tau(k) \propto k^{-2/3}$ (the Kolmogorov
scaling), the parameter $\sigma_{_{B}} = (4 / 135) \, (1 + 9
\epsilon)$. This case was considered by Rogachevskii and Kleeorin
(2003; 2004). When $\epsilon=0$ (there are no magnetic fluctuations
in the background turbulence due to the small-scale dynamo), the
shear-current dynamo occurs for $\mu < 1$. The boundary $\mu =1$
corresponds to the spatial scaling of the correlation time $\tau(k)
\propto k^{-1}$. For the Kolmogorov's type turbulence (i.e., for a
turbulence with a constant energy flux over the spectrum), the
energy spectrum $E(k) = - d\tau / dk \propto k^{-2}$. This implies
that the velocity is dominated by the large scales more strongly
than for the turbulence with a purely Kolmogorov spectrum $E(k)
\propto k^{-5/3}$.

For small hydrodynamic and magnetic Reynolds numbers, the turbulent
time $\tau(k) \propto 1/(\nu k^2)$ or $\tau(k) \propto 1/ (\eta
k^2)$ depending on the magnetic Prandtl number, i.e., $\tau(k)
\propto k^{-2}$. Then $\sigma_{_{B}} = (4 I_0 / 15) \, (- 1 + 3 \,
\epsilon) $, where $\mu = 2$, $\, \, \nu$ is the kinematic viscosity
and $\eta$ is the magnetic diffusivity due to electrical
conductivity of the fluid. When $\epsilon=0$ the parameter
$\sigma_{_{B}} < 0$, and there is no dynamo action due to the
shear-current effect in agreement with the recent studies by
R\"{a}dler and Stepanov (2006) and R\"{u}diger and Kichatinov
(2006). They have not found the dynamo action in nonrotating and
nonhelical shear flows with $\epsilon=0$ in the framework of the
second order correlation approximation (SOCA). This approximation is
valid for small hydrodynamic Reynolds numbers. Even in a highly
conductivity limit (large magnetic Reynolds numbers), SOCA can be
valid only for small Strouhal numbers, while for large hydrodynamic
Reynolds numbers (i.e., for fully developed turbulence), the
Strouhal number is 1. When $\epsilon> 1/3$ the mean magnetic field
can be generated due to the shear-current effect even for small
hydrodynamic and magnetic Reynolds numbers. However, the latter case
seems to be not realistic.

The effect of shear on the mean electromotive force and
shear-current effect have been studied by Rogachevskii and Kleeorin
(2003, 2004) for large hydrodynamic and magnetic Reynolds numbers
using two different approaches: the spectral $\tau$ approximation
(the third-order closure procedure) and the stochastic calculus,
i.e., the Feynman-Kac path integral representation of the solution
of the induction equation and Cameron-Martin-Girsanov theorem.

Note that Ruderman and Ruzmaikin (1984) formally constructed an
example of an exponentially growing magnetic field in a fluid with
shear and a homogeneous anisotropic magnetic diffusivity. An
essential condition for generation is that the vector defining the
anisotropy in this phenomenological model must be non-parallel and
non-perpendicular to the velocity. However, equations~(\ref{E2})
and~(\ref{E3}) are different from those given by Ruderman and
Ruzmaikin (1984) because they have not study the effect of shear on
the mean electromotive force. The first (although incorrect) attempt
to determine the effect of shear on the mean electromotive force has
been made by Urpin (1999a, 1999b) in the framework of SOCA.

In order to study the kinematic dynamo due to the shear current
effect, let us rewrite equations~(\ref{E2}) and~(\ref{E3}) for the
mean magnetic field in the dimensionless form
\begin{eqnarray}
{\partial A \over \partial t} &=& D\, B'_y + A'' \;,
\label{F11} \\
{\partial B_y \over \partial t} &=& - A' + B''_y \;,
\label{F12}
\end{eqnarray}
where the mean magnetic field is ${\bf B} = B_y(t,z) \, {\bf e}_y +
S_\ast^{-1} \bec{\nabla} {\bf \times} [A(z) \, {\bf e}_y]$, i.e.,
$B_x(t,z) = - S_\ast^{-1} \, A'(z)$, the parameter $S_\ast = S \,
L^2 / \eta_{_{T}}$ is the dimensionless shear number and $D= (l_0 \,
S_\ast / L)^2 \, \sigma_{_{B}}$ is the dynamo number. We consider
the following boundary conditions for a layer of the thickness $2L$
in the $z$ direction:
\begin{eqnarray}
B_y(t,|z|=1) = 0; \quad A'(t,|z|=1) = 0 \;,
\label{DF12}
\end{eqnarray}
i.e., ${\bf B}(t,|z|=1) = 0$. In dimensionless equations~(\ref{F11})
and~(\ref{F12}) the length is measured in units of $L$, the time is
measured in units of the turbulent magnetic diffusion time $L^{2} /
\eta_{_{T}}$, the mean magnetic field ${\bf B}$ is measured in units
of the equipartition field $B_{\rm eq} = \sqrt{4 \pi \rho} \, u_0 $
determined by the turbulent kinetic energy and the turbulent
magnetic diffusion coefficient is measured in units of the
characteristic value of $\eta_{_{T}} = l_0 \, u_{0} / 3 $. The
solution of equations~(\ref{F11}) and~(\ref{F12}) reads
\begin{eqnarray}
B_y(t,z) &=& B_0 \, \exp(\gamma \, t) \, \cos (K_z z + \varphi) \;,
\label{M5} \\
B_x(t,z) &=& {l_0 \over L} \, K_z \, \sqrt{\sigma_{_{B}}} \, B_0 \,
\exp(\gamma \, t) \, \cos (K_z z + \varphi). \label{M6}
\end{eqnarray}
The growth rate of the mean magnetic field in the dimensionless form
is given by $\gamma = \sqrt{D} \, K_z - K_z^2$. The wave vector
$K_z$ is measured in units of $L^{-1}$ and the growth rate $\gamma$
is measured in units of the inverse turbulent magnetic diffusion
time $\eta_{_{T}} / L^{2}$.

For the symmetric mode the angle $\varphi =\pi \, n$, the wave
number $K_z=(\pi / 2) (2m + 1)$ and the mean magnetic field is
generated when the dynamo number $D > D_{\rm cr} = (\pi^2/4) (2m +
1)^2$, where $n, m = 0, 1, 2, ... \,$. For this mode the mean
magnetic field is symmetric relative to the middle plane $z=0$. For
the antisymmetric mode the angle $\varphi =(\pi / 2) \, (2n+1)$ with
$n = 0, 1, 2, ...$, the wave number $K_z=\pi \, m$ and the magnetic
field is generated when the dynamo number $D > D_{\rm cr} = \pi^2 \,
m^2$, where $m = 1, 2, 3, ... \,$. Note that for the shear-current
dynamo, the ratio of the field components $B_x / B_y$ is small
[i.e., $B_x / B_y \sim (l_0 / L) \, K_z \, \sqrt{\sigma_{_{B}}} \ll
1$ when $K_z$ is not very large], see equations~(\ref{M5})
and~(\ref{M6}). This feature is similar to that for the $\alpha
\Omega$ dynamo, whereby the poloidal component of the mean magnetic
field is much smaller than the toroidal field. The maximum growth
rate of the mean magnetic field, $\gamma_{\rm max} = D / 4$, is
attained at $K_z = K_m = \sqrt{D} / 2$. This corresponds to the
characteristic scale of the mean magnetic field variations $L_B = 2
\pi L / K_m = (4 \pi /\sqrt{D}) \, L$.

\section*{4. Nonlinear dynamo due to the shear-current effect}

Kinematic dynamo models predict a field that grows without limit,
and they give no estimate of the magnitude for the generated
magnetic field. In order to find the magnitude of the field, the
nonlinear effects which limit the field growth must be taken into
account. The nonlinear theory of the shear-current effect was
developed by Rogachevskii and Kleeorin (2004, 2006).

\subsection*{4.1. The algebraic nonlinearity}

First, let us start with the algebraic nonlinearity which is
determined by the effects of the mean magnetic field on the motion
of fluid and on the cross-helicity. These effects cause quenching of
the mean electromotive force.

Below we outline the procedure of the derivation of the equations
for the nonlinear coefficients defining the mean electromotive force
in a homogeneous turbulence with a mean velocity shear (see for
details, Rogachevskii and Kleeorin 2004). We use the momentum
equation and the induction equation for the turbulent fields written
in a Fourier space. We derive equations for the correlation
functions of the velocity field $f_{ij}=\langle u_i u_j \rangle $,
the magnetic field $h_{ij}=\langle b_i b_j \rangle $ and the
cross-helicity $g_{ij}=\langle b_i u_j \rangle $, where the angular
brackets denote ensemble averaging. We split the tensors $ f_{ij},
h_{ij}$ and $g_{ij}$ into nonhelical, $h_{ij},$ and helical,
$h_{ij}^{(H)},$ parts. The helical part of the tensor $h_{ij}^{(H)}$
for magnetic fluctuations depends on the magnetic helicity, and it
is determined by the dynamic equation which follows from the
magnetic helicity conservation arguments (see Section 4.2). Then we
split the nonhelical parts of the correlation functions $ f_{ij},
h_{ij}$ and $g_{ij}$ into symmetric and antisymmetric tensors with
respect to the wave vector ${\bf k}$.

The second-moment equations include the first-order spatial
differential operators $\hat{\cal N}$  applied to the third-order
moments $M^{(III)}$. A problem arises how to close the system, i.e.,
how to express the set of the third-order terms $\hat{\cal N}
M^{(III)}$ through the lower moments $M^{(II)}$ (see, e.g., Orszag
1970; Monin and Yaglom 1975; McComb 1990). Various approximate
methods have been proposed in order to solve it. A widely used
spectral $\tau$ approximation (see, e.g., Orszag 1970, Pouquet {\it
et al.} 1976, Kleeorin {\it et al.} 1990, Kleeorin {\it et al.}
1996, Blackman and Field 2002, Rogachevskii and Kleeorin 2004,
Brandenburg {\it et al.} 2004, Brandenburg and Subramanian 2005a)
postulates that the deviations of the third-moment terms, $\hat{\cal
N} M^{(III)}({\bf k})$, from the contributions to these terms
afforded by the background turbulence, $\hat{\cal N}
M_0^{(III)}({\bf k})$, are expressed through the similar deviations
of the second moments, $M^{(II)}({\bf k}) - M_0^{(II)}({\bf k})$:
\begin{eqnarray}
\hat{\cal N} M^{(III)}({\bf k}) &-& \hat{\cal N} M_0^{(III)}({\bf
k})
\nonumber\\
&=& - {M^{(II)}({\bf k}) - M_0^{(II)}({\bf k}) \over \tau(k)} \;,
\label{TA1}
\end{eqnarray}
where $\tau(k)$ is the characteristic relaxation time, which can be
identified with the correlation time of the turbulent velocity
field. The background turbulence is determined by the budget
equations and the general structure of the moments is obtained by
symmetry reasoning.  In the background turbulence, the mean magnetic
field is zero. We applied the spectral $ \tau $-approximation only
for the nonhelical part $h_{ij}$ of the tensor of magnetic
fluctuations. We consider an intermediate nonlinearity which implies
that the mean magnetic field is not enough strong in order to affect
the correlation time of turbulent velocity field. The theory for a
very strong mean magnetic field can be corrected after taking into
account a dependence of the correlation time of the turbulent
velocity field on the mean magnetic field.

We assume that the characteristic time of variation of the mean
magnetic field ${\bf B}$ is substantially larger than the
correlation time $\tau(k)$ for all turbulence scales (which
corresponds to the mean-field approach). This allows us to get a
stationary solution for the equations for the second moments $
f_{ij}, \, h_{ij} $ and $g_{ij} $. For the integration in $ {\bf k}
$-space of these second moments we have to specify a model for the
background turbulence (with $ {\bf B} = 0)$. We use a simple model
for the background homogeneous and isotropic turbulence. Using the
derived equations for the second moments $ f_{ij}, \, h_{ij} $ and
$g_{ij} $ we calculate the mean electromotive force $ {\cal E}_{i} =
\varepsilon_{imn} \int g_{nm}({\bf k}) \,d {\bf k}$. This procedure
allows us to derive equations for the nonlinear coefficients
defining the mean electromotive force in a homogeneous turbulence
with a mean velocity shear (see for details, Rogachevskii and
Kleeorin 2004, 2006).

For simplicity in this study we do not take into account a quenching
of the turbulent magnetic diffusion. This facet is discussed in
details by Rogachevskii and Kleeorin (2004, 2006). We consider the
nonlinear dynamo problem with the algebraic nonlinearity
$\sigma_{_{N}}(B)$ which determines the nonlinear shear-current
effect. The mean magnetic field is determined by the following
nonlinear equations
\begin{eqnarray}
{\partial A \over \partial t} &=& D\, \sigma_{_{N}}(B) \, B'_y + A''
\;,
\label{NF11} \\
{\partial B_y \over \partial t} &=& - A' + B''_y \;, \label{NF12}
\end{eqnarray}
(see Rogachevskii and Kleeorin 2004, Rogachevskii {\it et al.}
2006), where $B=|{\bf B}|$. The function $\sigma_{_{N}}(B)$ defining
the nonlinear shear-current effect (which is normalized by
$\sigma_{_{B}})$, is shown in figure~5 for different values of the
parameter $\epsilon$. The asymptotic formulas for the nonlinear
function $\sigma_{_{N}}(B)$ are given by $\sigma_{_{N}}(B) = 1$ for
a weak mean magnetic field $(B \ll B_{\rm eq} /4)$ and
$\sigma_{_{N}}(B) = - 11 (1 + \epsilon) / 4 (1 + 9\epsilon)$ for $B
\gg B_{\rm eq} / 4$ (see figure~5). This implies that the nonlinear
function $\sigma_{_{N}}(B)$ defining the shear-current effect
changes its sign at some value of the mean magnetic field
$B=B_\ast$. Here $B_\ast = 1.2 B_{\rm eq}$ for $\epsilon=0$ and
$B_\ast = 1.4 B_{\rm eq}$ for $\epsilon=1$. However, there is no
quenching of the nonlinear shear-current effect contrary to the
quenching of the nonlinear alpha effect, the nonlinear turbulent
magnetic diffusion, etc. The background magnetic fluctuations
(caused by the small-scale dynamo and described by the parameter
$\epsilon )$, affect the nonlinear function $\sigma_{_{N}}({\bf
B})$.

\begin{figure}
\vspace*{2mm} \centering
\includegraphics[width=8cm]{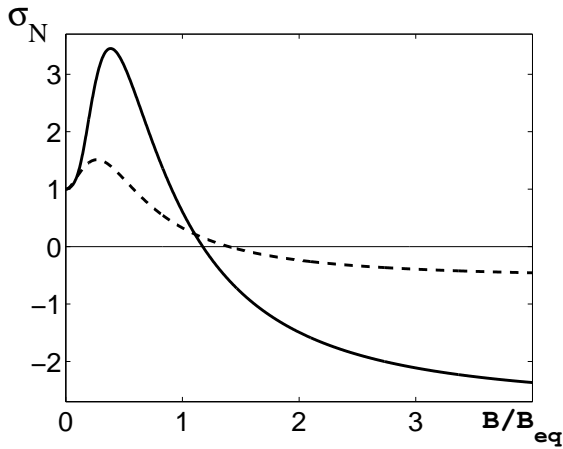}
\caption{\label{Fig5} The dimensionless nonlinear coefficient
$\sigma_{_{N}}(B)$ defining the shear-current effect for different
values of the parameter $\epsilon$: $\, \, \, \epsilon=0$ (solid);
$\epsilon=1$ (dashed).}
\end{figure}

Numerical solutions of equations~(\ref{NF11}) and~(\ref{NF12}) for
the nonlinear problem with the algebraic nonlinearity
$\sigma_{_{N}}({\bf B})$ are plotted in figures~6-8. In particular,
these figures show the nonlinear evolution of the mean magnetic
field $B(t,z=0)$ due to the shear-current effect for different
values of the dynamo number $D$ and the parameter $\epsilon$. The
magnitude of the saturated mean magnetic field is in several times
larger than the equipartition field depending on the dynamo number.
Inspection of figures~7-8 shows that there is a range in the dynamo
number $D=22.8 - 59$ where the nonlinear oscillations of mean
magnetic field are observed at $\epsilon=0$.

\begin{figure}
\vspace*{2mm} \centering
\includegraphics[width=8cm]{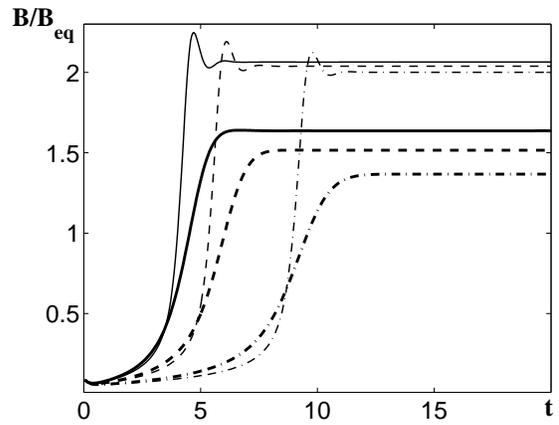}
\caption{\label{Fig6} The nonlinear evolution of the mean magnetic
field $B(t,z=0)$ due to the shear-current effect with the algebraic
nonlinearity for $\epsilon=0$ (thin curve) and $\epsilon=1$ (thick
curve) and different near-threshold values of the dynamo number: $\,
\, \, D=1.45 \, D_{\rm cr}$ (solid); $D=1.3 \, D_{\rm cr}$ (dashed);
$D=1.15 \, D_{\rm cr}$ (dashed-dotted).}
\end{figure}

\begin{figure}
\vspace*{2mm} \centering
\includegraphics[width=8cm]{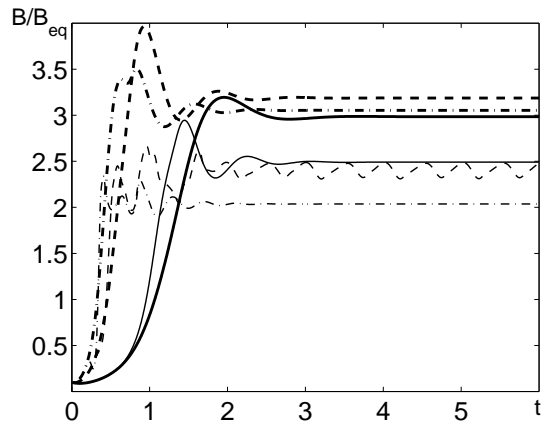}
\caption{\label{Fig7} The nonlinear evolution of the mean magnetic
field $B(t,z=0)$ due to the shear-current effect with the algebraic
nonlinearity for $\epsilon=0$ (thin curve) and $\epsilon=1$ (thick
curve) and different values of the dynamo number: $\, \, \, D=10$
(solid); $D=30$ (dashed); $D=50$ (dashed-dotted).}
\end{figure}

\begin{figure}
\vspace*{2mm} \centering
\includegraphics[width=8cm]{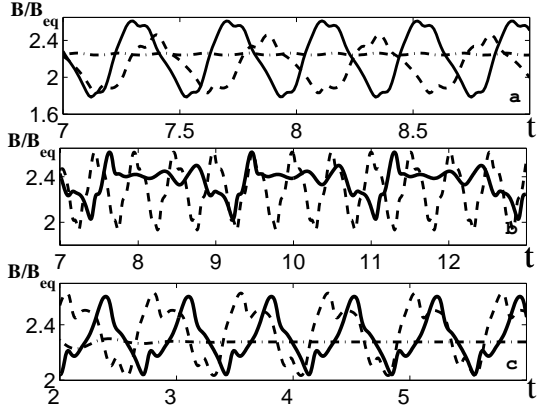}
\caption{\label{Fig8} The nonlinear evolution of the mean magnetic
field $B(t,z=0)$ due to the shear-current effect with the algebraic
nonlinearity for $\epsilon=0$ and different values of the dynamo
number: (a). $D=47$ (solid), $D=52$ (dashed), $D=59$
(dashed-dotted); (b). $D=31$ (solid), $D=35$ (dashed); (c). $D=28$
(solid); $D=22.77$ (dashed); $D = 22.765$ (dashed-dotted).}
\end{figure}

\section*{4.2. The algebraic and dynamic nonlinearities}

In this study we consider nonhelical and nonrotating homogeneous
turbulence. This implies that the kinetic helicity and the
hydrodynamic $\alpha$ effect are zero. However, the magnetic
$\alpha$ effect caused by the small-scale magnetic helicity is not
zero even in nonhelical turbulence. In particular, the magnetic
helicity conservation implies the growth of a magnetic alpha effect
independent of whether kinetic helicity is driven into the system.
In subsection 4.1 we have concentrated on the nonlinear
shear-current effect (the algebraic nonlinearity) and have not
discussed the effect of  small-scale magnetic helicity (the dynamic
nonlinearity) on the nonlinear saturation of the mean magnetic
field. In this subsection we study joint action of the algebraic and
dynamic nonlinearities.

The small-scale magnetic helicity causes the magnetic $\alpha$
effect which is given by $\alpha_m = \Phi_{_{N}}(B) \, \chi_c({\bf
B})$, where $\Phi_{_{N}}(B)$ is the quenching function of the
magnetic $\alpha$ effect given below by Eq.~(\ref{RM4}). The
function $\chi_{c}({\bf B}) \equiv (\tau_0 / 12 \pi \rho) \langle
{\bf b} {\bf \cdot} (\bec{\nabla} {\bf \times} {\bf b}) \rangle $ is
related to the small-scale current helicity $\langle {\bf b} {\bf
\cdot} (\bec{\nabla} {\bf \times} {\bf b}) \rangle$. For a weakly
inhomogeneous turbulence $(l_0 \ll L)$, the function $\chi_{c}$ is
proportional to the small-scale magnetic helicity, $\chi_{c} =
\chi_{m} / (18 \pi \eta_{_{T}} \rho)$ (see Kleeorin and Rogachevskii
1999), where $\chi_{m} = \langle {\bf a} {\bf \cdot} {\bf b}
\rangle$ is the small-scale magnetic helicity and ${\bf a}$ is the
vector potential of small-scale magnetic field. The function
$\chi_c({\bf B})$ entering the magnetic $\alpha$ effect is
determined by the following dynamical dimensionless equation (which
is derived using arguments based on the magnetic helicity
conservation law):
\begin{eqnarray}
{\partial \chi_{c} \over \partial t} + \bec{\nabla} {\bf \cdot} {\bf
F} + {\chi_{c} \over \tau_\chi} = - \biggl({2 L \over l_0} \biggr)^2
(\bec{\cal E} {\bf \cdot} {\bf B}) \;, \label{M30}
\end{eqnarray}
(see, e.g., Kleeorin and Ruzmaikin 1982, Gruzinov and Diamond 1994,
1996, Kleeorin {\it et al.} 1995, 2000, 2002, 2003a, 2003b, Kleeorin
and Rogachevskii 1999, Blackman and Field 2000, Blackman and
Brandenburg 2002, Brandenburg and Subramanian 2005a, Zhang {\it et
al.} 2006), where $\tau_\chi = (1/3) (l_0/L)^{2} {\rm Rm}$ is the
characteristic relaxation time of the small-scale magnetic helicity,
${\rm Rm}$ is the magnetic Reynolds number and ${\bf F}$ is related
to the flux of the small-scale magnetic helicity.

The simplest form of the magnetic helicity flux is the turbulent
diffusive flux of the magnetic helicity, ${\bf F} = - \kappa_{_{T}}
\bec{\nabla} \chi_c$ (see Kleeorin {\it et al.} 2002, 2003b), where
the turbulent diffusivity coefficient $\kappa_{_{T}}$ is measured in
units of $\eta_{_{T}}$ and the function $\chi_c$ is measured in
units of $\eta_{_{T}} / L$. In real systems, the flux of small-scale
magnetic helicity can be accompanied by some flux of large-scale
magnetic helicity. However, the flux of large-scale  magnetic
helicity does not explicitly enters in the dynamical
equation~(\ref{M30}) for the evolution of the small-scale  magnetic
helicity. This flux mostly affects the large-scale magnetic
helicity. It can also introduce an additional anisotropy of
turbulence, which can affect the dynamics of the mean magnetic
field.

Equation~(\ref{M30}) determines the dynamics of the small-scale
magnetic helicity, i.e. its production, dissipation and transport.
For very large magnetic Reynolds numbers (which are typical for many
astrophysical situations), the relaxation term $\chi_{c} /
\tau_\chi$ is very small, and it is very often dropped in
Eq.~(\ref{M30}) in spite of the fact that the small yet finite
magnetic diffusion is required for the reconnection of magnetic
field lines. In particular, the magnetic Reynolds number, ${\rm Rm}$
does not enter into the steady state solution of Eq.~(\ref{M30}) in
the limit of very large ${\rm Rm}$ due to the effect of the magnetic
helicity flux (see Kleeorin {\it et al.} 2003b).

The account for the dynamics of the small-scale magnetic helicity
results in that the mean magnetic field is determined by the
following dimensionless equations
\begin{eqnarray}
&& {\partial A(t,z) \over \partial t} = D \, \sigma_{_{N}}(B) \,
B'_y + B_y \, \Phi_{_{N}}(B) \, \chi_c(t,z) + A'' \;,
\nonumber\\
\label{RM1} \\
&& {\partial B_y(t,z) \over \partial t} = - A' + B''_y \;,
\label{RM2} \\
&& {\partial \chi_c(t,z) \over \partial t} - \kappa_{_{T}} \,
\chi_c'' + {\chi_{c} \over \tau_\chi} = C \, \biggl(A' \, B'_y - B_y
\, [D \, \sigma_{_{N}}(B) \, B'_y
\nonumber\\
&& \quad \quad \quad \quad + B_y \, \Phi_{_{N}}(B) \, \chi_c(t,z)  +
A''] \biggr) \;, \label{RM3}
\end{eqnarray}
where $\alpha_m = S_\ast^{-1} \, \Phi_{_{N}}(B) \, \chi_c(t,z) $ is
the magnetic $\alpha$ effect and the parameter $C = (2 L / l_0)^2$.
In equations~(\ref{RM2}) and~(\ref{RM3}) we have neglected small
terms $\sim {\rm O}(S_\ast^{-2}) \ll 1$. The algebraic function
$\Phi_{_{N}}(B)$ in these equations is given by
\begin{eqnarray}
\Phi_{_{N}}(B) = {3 \over 8 B^2} \biggl[1 - {\arctan (\sqrt{8} B)
\over \sqrt{8} B} \biggr] \;, \label{RM4}
\end{eqnarray}
(see, e.g., Rogachevskii and Kleeorin 2000), where $\Phi_{_{N}}(B) =
1 - (24/5) B^2$ for $B \ll 1/\sqrt{8} $ and $\Phi_{_{N}}(B) =
3/(8B^2)$ for $B \gg 1/\sqrt{8} $. Here the mean magnetic field $B$
is measured in units of the equipartition field $B_{\rm eq}$
determined by the turbulent kinetic energy. In
equations~(\ref{RM1})-(\ref{RM3}) there are four parameters: the
dynamo number $D = 4 \, S_\ast^2 \, \sigma_{_{B}} / C$, the
turbulent diffusivity of the small-scale magnetic helicity
$\kappa_{_{T}}$ (measured in units of $\eta_{_{T}}$), the parameter
$C = (2 L / l_0)^2$ and the relaxation time of the small-scale
magnetic helicity $\tau_\chi = (4/3) {\rm Rm} / C$. Consider the
simple boundary conditions for a layer of the thickness $2L$ in the
$z$ direction: $B_y(t,|z|=1) = 0$, $\,\, A'(t,|z|=1) = 0$ and
$\chi_c(t,|z|=1) = 0$, where $z$ is measured in units of $L$. The
initial conditions for the symmetric mode are chosen in the form:
$B_y(t=0,z) = B_0 \, \cos (\pi z / 2)$, $\, \, A(t=0,z) = 0$ and
$\chi_c(t=0,z) = 0$.

\begin{figure}
\vspace*{2mm} \centering
\includegraphics[width=8cm]{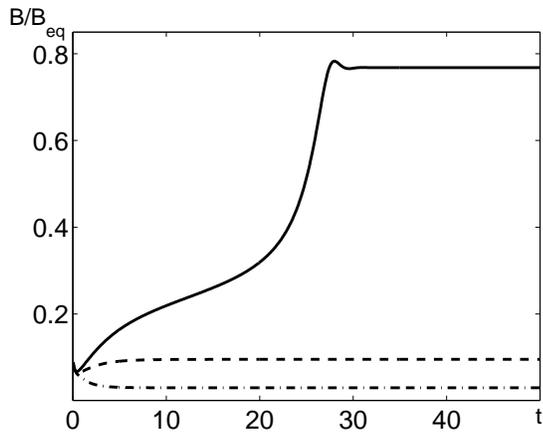}
\caption{\label{Fig9} The nonlinear evolution of the mean magnetic
field $B(t,z=0)$ due to the shear-current effect with the algebraic
and dynamic nonlinearities for $\epsilon=0$, $\, \, D=2 \, D_{\rm
cr}$, $\, \, C = 100$ and different values of the parameter
$\kappa_{_{T}}$:  $\, \, \, \kappa_{_{T}} = 0.5$ (solid); $\, \,
\kappa_{_{T}} = 0.3$ (dashed); $\, \, \kappa_{_{T}} = 0.1$
(dashed-dotted).}
\end{figure}

\begin{figure}
\vspace*{2mm} \centering
\includegraphics[width=8cm]{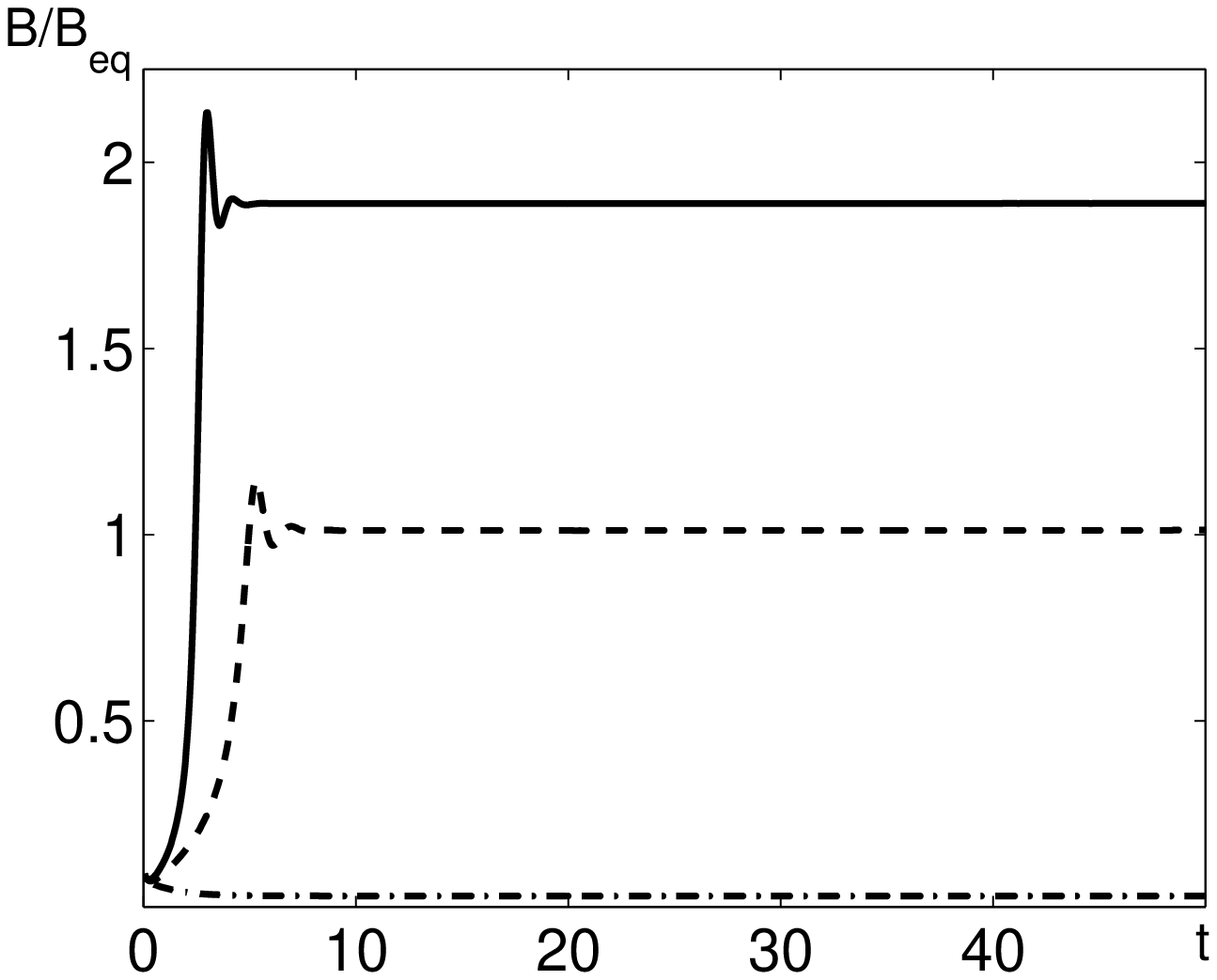}
\caption{\label{Fig10} The nonlinear evolution of the mean magnetic
field $B(t,z=0)$ due to the shear-current effect with the algebraic
and dynamic nonlinearities for $\epsilon=0$, $\, \, D=2 \, D_{\rm
cr}$, $\, \, C = 100$ and different values of the parameter
$\kappa_{_{T}}$:  $\, \, \, \kappa_{_{T}} = 10$ (solid); $\, \,
\kappa_{_{T}} = 1$ (dashed); $\, \, \kappa_{_{T}} = 0.1$
(dashed-dotted).}
\end{figure}

\begin{figure}
\vspace*{2mm} \centering
\includegraphics[width=8cm]{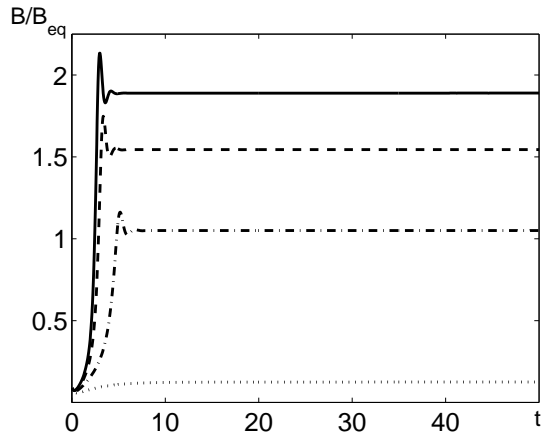}
\caption{\label{Fig11} The nonlinear evolution of the mean magnetic
field $B(t,z=0)$ due to the shear-current effect with the algebraic
and dynamic nonlinearities for $\epsilon=0$, $\, \, \kappa_{_{T}} =
10$, $\, \, D=2 \, D_{\rm cr}$ and different values of the parameter
$C$: $\, \, \, C = 100$ (solid); $\, \, C = 300$ (dashed); $\, \, C
= 900$ (dashed-dotted); $\, \, C = 2700$ (dotted).}
\end{figure}

Numerical solutions of equations~(\ref{RM1})-(\ref{RM3}) for the
nonlinear problem with the algebraic and dynamic nonlinearities are
plotted in figures~9-13. In particular, the nonlinear evolution of
the mean magnetic field $B(t,z=0)$ for different values of the
parameters $\kappa_{_{T}}$, $\, C$, the dynamo numbers $D$ and very
large magnetic Reynolds numbers ${\rm Rm}$ is shown in figures~9-13.
Inspection of figures~9-10 shows that the saturated level of the
mean magnetic field depends strongly on the value of the turbulent
diffusivity of the magnetic helicity $\kappa_{_{T}}$. The saturated
level of the mean magnetic field changes from very small value for
$\kappa_{_{T}} = 0.1$ to the super-equipartition field for
$\kappa_{_{T}} = 10$. This is an indication of very important role
of the transport of the magnetic helicity for the saturated level of
the mean magnetic field. Indeed, during the generation of the mean
magnetic field, the product $\bec{\cal E} {\bf \cdot} {\bf B}$ is
positive, and this produces negative contribution to the small-scale
magnetic helicity (and negative magnetic $\alpha$ effect, see
Eq.~(\ref{M30})). Therefore, this reduces the rate of generation of
large-scale magnetic field because the first and the second terms in
the right hand side of Eq.~(\ref{RM1}) have opposite signs. The
first term in Eq.~(\ref{RM1}) describes the shear-current effect,
while the second term in Eq.~(\ref{RM1}) determines the magnetic
$\alpha$ effect. If the magnetic helicity does not effectively
transported out from the generation region, the mean magnetic field
is saturated even at small value of the magnetic field. Increase of
the magnetic helicity flux (by increasing of turbulent diffusivity
$\kappa_{_{T}}$ of magnetic helicity) results in increase of the
saturated level of the mean magnetic field above the equipartition
field (see figures~9-10). Note that increase of the parameter $C$
decreases the saturated level of the mean magnetic field (see
figure~11). Actually the ratio $\kappa_{_{T}} /C$ determines the
saturated level of the mean magnetic field in a steady-state.

\begin{table}
\label{tab1}
\begin{tabular}{|l|c|c|}
\multicolumn{3}{c}{Table 1}\\
\multicolumn{3}{c}{The saturated mean magnetic field}\\
\multicolumn{3}{c}{versus the magnetic Reynolds number}\\
\hline
       & $\kappa_{_{T}}=0.1$     & $\kappa_{_{T}}=0.3$
 \\
 \hline
 ${\rm Rm}$  & $B / B_{\rm eq}$     & $B / B_{\rm eq}$
 \\
 \hline
 7.5  & 1.16  & 1.2 \\
 \hline
 15  & 0.89  & 1\\
\hline
 16.5  & 0.37  & 0.97 \\
 \hline
  17  & 0.2 & 0.967 \\
 \hline
 30  & 0.1  & 0.83 \\
 \hline
 36   &  0.085 & 0.35    \\
 \hline
 50   & 0.076   & 0.16 \\
 \hline
 $10^2$  & 0.05  & 0.12 \\
 \hline
  $10^6$ & 0.03 & 0.096 \\
 \hline

\end{tabular}
\end{table}

\begin{figure}
\vspace*{2mm} \centering
\includegraphics[width=8cm]{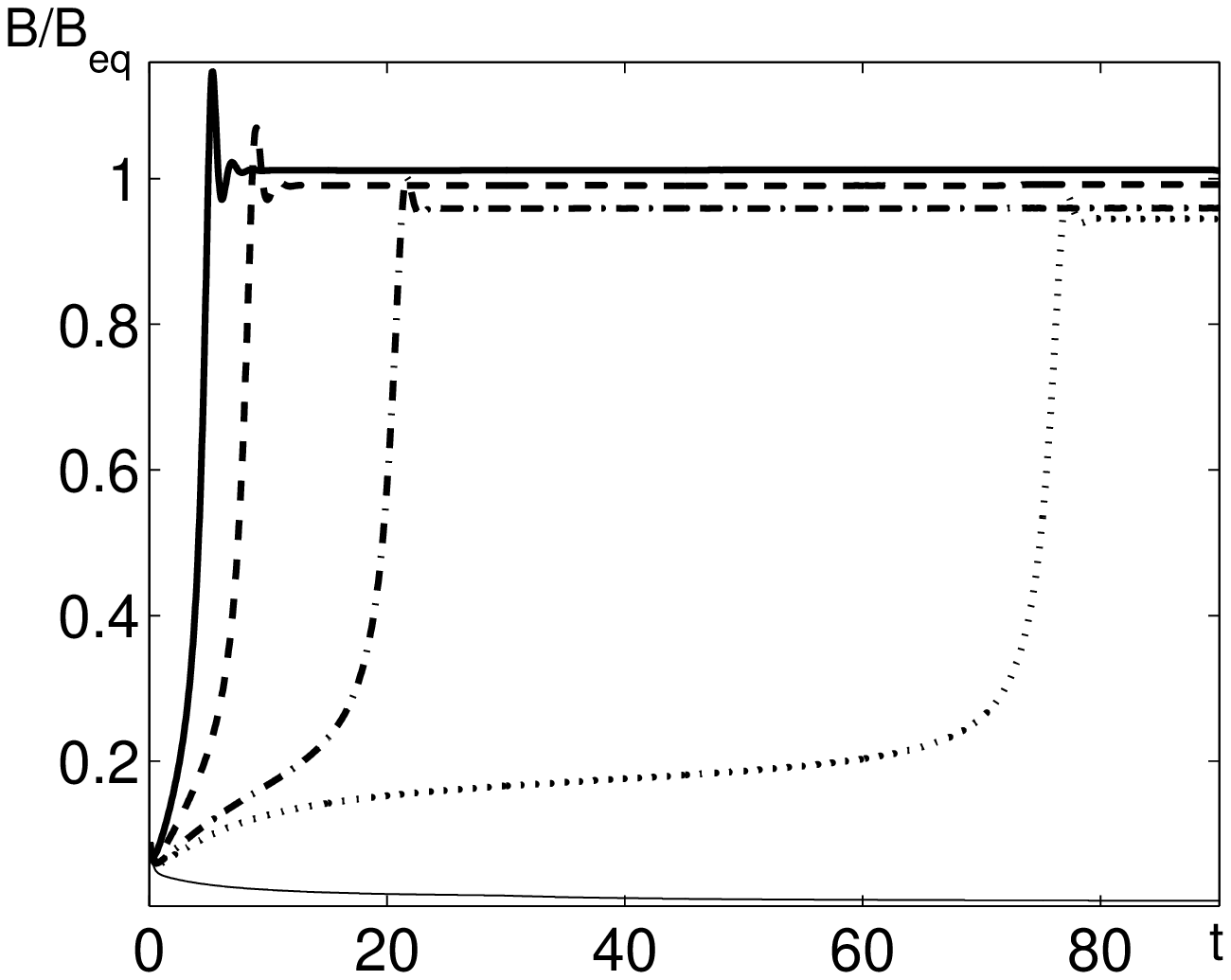}
\caption{\label{Fig12} The nonlinear evolution of the mean magnetic
field $B(t,z=0)$ due to the shear-current effect with the algebraic
and dynamic nonlinearities for $\epsilon=0$, $\, \, \kappa_{_{T}} =
1$, $\, \, C = 100$ and different values of the dynamo number: (a).
$D=2 \, D_{\rm cr}$ (thick solid); $\, \, D=1.7 \, D_{\rm cr}$
(dashed); $D=1.5 \, D_{\rm cr}$ (dashed-dotted); $D=1.43 \, D_{\rm
cr}$ (dotted); $D=1.1 \, D_{\rm cr}$ (thin solid).}
\end{figure}

\begin{figure}
\vspace*{2mm} \centering
\includegraphics[width=8cm]{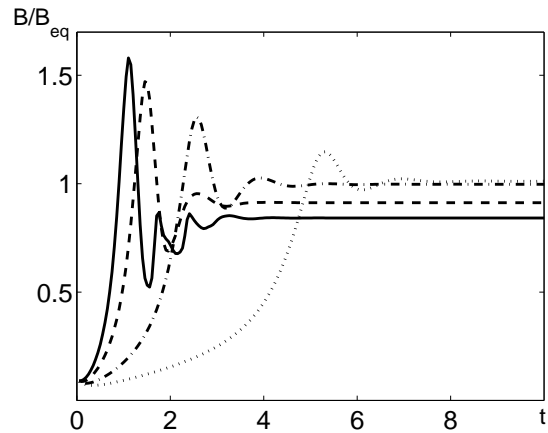}
\caption{\label{Fig13} The nonlinear evolution of the mean magnetic
field $B(t,z=0)$ due to the shear-current effect with the algebraic
and dynamic nonlinearities for $\epsilon=0$, $\, \, \kappa_{_{T}} =
1$, $\, \, C = 100$ and different values of the dynamo number: (a).
$D=7 \, D_{\rm cr}$ (solid); $\, \, D=5 \, D_{\rm cr}$ (dashed);
$D=3 \, D_{\rm cr}$ (dashed-dotted); $D=2 \, D_{\rm cr}$ (dotted).}
\end{figure}

For the cases shown in figures~9-13, we drop the small relaxation
term $\chi_{c} / \tau_\chi$ in Eq.~(\ref{M30}) due to very large
magnetic Reynolds numbers. Now we consider moderate magnetic
Reynolds numbers, when the relaxation term $\chi_{c} / \tau_\chi$ in
Eq.~(\ref{M30}) is not small but the flux of magnetic helicity is
weak (e.g., $\kappa_{_{T}} = 0.1 - 0.3)$. In this case the
small-scale magnetic helicity does not effectively transported out
from the generation region by the helicity flux. In Table~1 we
demonstrate the effect of the moderate magnetic Reynolds numbers on
the saturated level of the mean magnetic field. The decrease of the
magnetic Reynolds numbers (and the relaxation time $\tau_\chi)$
increases the saturated level of the mean magnetic field (see
Table~1), because the relaxation term $\chi_{c} / \tau_\chi$  in
Eq.~(\ref{M30}) decreases the small-scale magnetic helicity in the
generation region. On the other hand, for larger flux of small-scale
magnetic helicity $(\kappa_{_{T}} \geq 0.5)$, the effect of the
magnetic Reynolds numbers on the saturated level of the mean
magnetic field is very small. Note that the moderate magnetic
Reynolds numbers ${\rm Rm} = 10 - 50$ are irrelevant for
astrophysical applications, although they are of an interest for the
direct numerical simulations.

Figures~12-13 show the nonlinear evolution of the mean magnetic
field $B(t,z=0)$ for different values of the dynamo numbers $D$. The
saturated level of the mean magnetic field increases with the
increase of the dynamo numbers $D$ within the range $D_{\rm cr} < D
< 2 \, D_{\rm cr}$, and it decreases with the increase of the dynamo
number for $D > 2 \, D_{\rm cr}$. This is a new feature in the
nonlinear mean-field dynamo. For example, in the $\alpha\Omega$
dynamo the saturated level of the mean magnetic field usually
increases with the increase the dynamo numbers.

Generation of the large-scale magnetic field in a nonhelical
turbulence with an imposed mean velocity shear has been recently
investigated by Brandenburg (2005) and Brandenburg {\it et al.}
(2005) using direct numerical simulations. The results of these
numerical simulations are in a good agreement with the numerical
solutions of the nonlinear dynamo equations~(\ref{RM1})-(\ref{RM3})
discussed in section 4.

Now let us compare the results of the numerical solutions of the
nonlinear dynamo equations~(\ref{RM1})-(\ref{RM3}) with the
numerical study by Brandenburg and Subramanian (2005b) of the
mean-field dynamo with large-scale shear. In our study we use the
expression for the nonlinear electromotive force determined by
Rogachevskii and Kleeorin (2004, 2006), which includes the nonlinear
shear-current effect. On the other hand, Brandenburg and Subramanian
(2005b) use very simplified form of the mean electromotive force,
neglecting e.g., the $\bec{\kappa}$ effect related to the symmetric
parts of the gradient tensor of the mean magnetic field. This effect
contributes to the shear-current effect (see Rogachevskii and
Kleeorin  2003, 2004). Brandenburg and Subramanian (2005b) have not
taken into account the properties of the  nonlinear shear-current
effect found by Rogachevskii and Kleeorin (2004, 2006). In
particular, there is no quenching of the nonlinear shear-current
effect contrary to the quenching of the nonlinear alpha effect, the
nonlinear turbulent magnetic diffusion, etc. During the nonlinear
growth of the mean magnetic field, the shear-current effect only
changes its sign at some value of the mean magnetic field which
affects the level of the saturated mean magnetic field (see
Rogachevskii and Kleeorin 2004, 2006). In our study we neglect small
terms $\sim {\rm O}(S_\ast^{-2}) \ll 1$ in
equations~(\ref{RM1})-(\ref{RM3}), i.e., we do not consider
$\alpha_m^2$ effect because the parameter $S_\ast$ should be very
large (see section 3). In addition we do not consider $\delta^2 \,
S$ effect because we neglected the small terms $\sim {\rm
O}[(l_0/L)^2]$ in equation~(\ref{RM2}). Here the parameter $\delta$
determines the ${\bf W} {\bf \times} {\bf J}$ term in the mean
electromotive force. In our numerical solutions of the nonlinear
mean-field dynamo equations we use the simplest form of the magnetic
helicity flux (i.e., we use the turbulent diffusive flux of the
magnetic helicity, ${\bf F} = - \kappa_{_{T}} \bec{\nabla} \chi_c$,
where $\kappa_{_{T}}$ is not small), while Brandenburg and
Subramanian (2005b) use the current helicity flux of Vishniac and
Cho (2001) and a very small turbulent diffusive flux of the magnetic
helicity (with $\kappa_{_{T}}= 10^{-2})$.

The parameter range in the study by Brandenburg and Subramanian
(2005b) is different from that used in our study, and the maximum
saturated level of the mean magnetic field obtained in the study by
Brandenburg and Subramanian (2005b) is $B= 0.6 B_{\rm eq}$ at ${\rm
Rm} = 10^2$, which strongly decreases with increase the magnetic
Reynolds number ${\rm Rm}$ (see Table 4 of Brandenburg and
Subramanian 2005b). On the other hand, in our numerical solutions of
the nonlinear mean-field dynamo equations which take into account
the shear-current effect, the saturated level of the mean magnetic
field reaches the super-equipartition field. These are the reasons
why our numerical results discussed in this section are different
from those obtained by Brandenburg and Subramanian (2005b). Note
however, that increase of the saturated level of the mean magnetic
field with the decrease of the magnetic Reynolds numbers in the case
of very weak flux of small-scale magnetic helicity (found by
Brandenburg and Subramanian 2005b), is confirmed by our numerical
study (see Table~1).

\section*{5. Discussion}

In this study we show that in a sheared nonhelical homogeneous
turbulence the large-scale magnetic field can grow due to the
shear-current effect from a very small seeding magnetic field. The
shear-current dynamo strongly depends on the spatial scaling of the
correlation time $\tau(k)$ of the background turbulence. In
particular, for Kolmogorov scaling, $\tau(k) \propto k^{-2/3}$, the
dynamo instability due to the shear-current effect occurs, while
when $\tau(k) \propto k^{-2}$ (for small hydrodynamic and magnetic
Reynolds numbers) there is no the dynamo action in a sheared
nonhelical turbulence. The dynamo instability is saturated by the
nonlinear effects, and the dynamical nonlinearity due to the
evolution of small-scale magnetic helicity, plays a crucial role in
nonlinear saturation of the large-scale magnetic field. The magnetic
helicity flux strongly affects the saturated level of the mean
magnetic field in the nonlinear stage of the dynamo action. In
particular, our numerical solutions of the nonlinear mean-field
dynamo equations which take into account the shear-current effect,
show that if the magnetic helicity flux is not small, the saturated
level of the mean magnetic field is of the order of the
equipartition field determined by the turbulent kinetic energy.

The shear-current dynamo acts also in inhomogeneous turbulence.
However, in inhomogeneous turbulence with a large-scale velocity
shear the kinetic helicity and hydrodynamic $\alpha$ effect are not
zero (see Rogachevskii and Kleeorin 2003; 2006; R\"{a}dler and
Stepanov 2006). In this case the shear-current dynamo acts together
with the $\alpha$-shear dynamo (which is similar to the $\alpha
\Omega$ dynamo). The joint action of the shear-current and the
$\alpha$-shear dynamo has been recently discussed by Rogachevskii
and Kleeorin (2006); Pipin (2006).

Turbulence with a large-scale velocity shear is a universal feature
in astrophysics, and the obtained results can be important for
explanation of the large-scale magnetic fields generated in
astrophysical sheared turbulence. Rogachevskii {\it et al.} (2006)
have suggested that the shear-current effect might be considered as
an origin for the large-scale magnetic fields in colliding
protogalactic clouds and in merging protostellar clouds.

Note that the problem of the generation of the mean magnetic field
in a turbulence with large-scale velocity shear is similar to that
for generation of mean vorticity in a sheared turbulence. The
instability of the perturbations of the mean vorticity in a
turbulence with a large-scale linear velocity shear was studied by
Elperin {\it et al.} (2003). This instability is caused by a
combined effect of the large-scale shear motions (skew-induced
deflection of equilibrium mean vorticity due to the shear) and
Reynolds-stress-induced generation of perturbations of mean
vorticity. In Appendix~A we compare the problem of generation of the
mean magnetic field in a turbulence with a large-scale velocity
shear with that of the generation of mean vorticity in a sheared
turbulence.

\section*{Acknowledgments}

We have benefited from stimulating suggestions made by Alexander
Schekochihin. I.R. thanks for hospitality the Department of Applied
Mathematics and Theoretical Physics of University of Cambridge.

\bigskip
\leftline{\bf References}
\bigskip
{
\parindent 0em
  \hangindent 2em
  \parskip2ex

    Beck, R., Magnetic fields in the Milky Way and other spiral
galaxies. In {\it How Does the Galaxy Work?}, edited by E.J. Alfaro,
E. P\'erez and J. Franco, pp.277-286, 2004, Kluwer, Dordrecht.

    Beck, R., Magnetic fields in galaxies. In {\it Cosmic Magnetic Fields},
edited by R. Wielebinski and R. Beck, pp.41-68, 2005, Springer,
Berlin.

    Blackman, E.G. and Brandenburg, A., Dynamic nonlinearity in large
scale dynamos with shear. {\it Astrophys. J.}, 2002, {\bf 579},
359-373.

    Blackman, E.G. and Field, G., Constraints on the magnitude of alpha
in dynamo theory. {\it Astrophys. J.}, 2000, {\bf 534}, 984-988.

    Blackman, E.G. and Field, G., New dynamical mean-field dynamo
theory and closure approach. {\it Phys. Rev. Lett.}, 2002, {\bf 89},
265007 (1-4).

    Brandenburg, A., The case for a distributed solar dynamo
shaped by near-surface shear. {\it Astrophys. J.}, 2005, {\bf 625},
539-547.

    Brandenburg, A., Haugen, N.E.L., K\"{a}pyl\"{a}, P.J. and Sandin, C.,
The problem of small and large scale fields in the solar dynamo.
{\it Astron. Nachr.}, 2005, {\bf 326}, 174-185.

    Brandenburg, A., K\"{a}pyl\"{a}, P. and Mohammed, A., Non-Fickian
diffusion and tau-approximation from numerical turbulence. {\it
Phys. Fluids}, 2004, {\bf 16}, 1020-1027.

    Brandenburg, A. and Subramanian, K., Astrophysical magnetic fields
and nonlinear dynamo theory. {\it Phys. Rept.}, 2005a,  {\bf 417},
1-209.

    Brandenburg, A. and Subramanian, K., Strong mean field
dynamos require supercritical helicity fluxes. {\it Astron. Nachr.},
2005b, {\bf 326}, 400-408.

    Elperin, T., Kleeorin, N. and Rogachevskii I., Generation of
large-scale vorticity in a homogeneous turbulence with a mean
velocity shear. {\it Phys. Rev. E}, 2003, {\bf 68}, 016311 (1-8).

    Gruzinov, A.  and Diamond, P.H., Self-consistent theory of
mean-field electrodynamics. {\it Phys. Rev. Lett.}, 1994, {\bf 72},
1651-1653.

    Gruzinov, A.  and Diamond, P.H., Nonlinear mean-field
electrodynamics of turbulent dynamos. {\it Phys. Plasmas}, 1996,
{\bf 3}, 1853-1857.

    Kleeorin N., Kuzanyan, K., Moss D., Rogachevskii I., Sokoloff D. and
Zhang, H., Magnetic helicity evolution during the solar activity
cycle: observations and dynamo theory. {\it Astron. Astrophys.},
2003a, {\bf 409}, 1097-1105.

    Kleeorin, N., Mond, M. and Rogachevskii, I., Magnetohydrodynamic
turbulence in the solar convective zone as a source of oscillations
and sunspots formation. {\it Astron. Astrophys.}, 1996, {\bf 307},
293-309.

    Kleeorin N., Moss D., Rogachevskii I. and Sokoloff D., Helicity
balance and steady-state strength of dynamo generated galactic
magnetic field. {\it Astron. Astrophys.}, 2000, {\bf 361}, L5-L8.

    Kleeorin N., Moss D., Rogachevskii I. and Sokoloff D., The role of
magnetic helicity transport in nonlinear galactic dynamos. {\it
Astron. Astrophys.}, 2002, {\bf 387}, 453-462.

    Kleeorin N., Moss D., Rogachevskii I. and Sokoloff D., Nonlinear
magnetic diffusion and magnetic helicity transport in galactic
dynamos. {\it Astron. Astrophys.}, 2003b, {\bf 400}, 9-18.

    Kleeorin, N. and Rogachevskii, I., Magnetic helicity tensor for an
anisotropic turbulence. {\it Phys. Rev. E}, 1999, {\bf 59},
6724-6729.

    Kleeorin, N., Rogachevskii, I. and Ruzmaikin, A., Magnetic force
reversal and instability in a plasma with advanced
magnetohydrodynamic turbulence. {\it Sov. Phys. JETP}, 1990, {\bf
70}, 878-883. Translation from {\it Zh. Eksp. Teor. Fiz.}, 1990,
{\bf 97}, 1555-1565.

    Kleeorin, N., Rogachevskii, I. and Ruzmaikin, A., Magnitude of
dynamo - generated magnetic field in solar - type convective zones.
{\it Astron. Astrophys.}, 1995, {\bf 297}, 159-167.

    Kleeorin, N. and Ruzmaikin, A., Dynamics of the averaged turbulent
helicity in a magnetic field. {\it Magnetohydrodynamics}, 1982, {\bf
18}, 116-122. Translation from {\it Magnitnaya Gidrodinamika}, 1982,
{\bf 2}, 17-24.

    Krause, F. and R\"{a}dler, K.-H., {\it Mean-Field
Magnetohydrodynamics and  Dynamo Theory}, 1980 (Pergamon: Oxford).

    McComb, W.D., {\it The Physics of Fluid Turbulence}, 1990
(Clarendon: Oxford).

    Moffatt, H.K., {\it Magnetic Field Generation in Electrically
Conducting  Fluids}, 1978 (Cambridge University Press: New York).

    Monin, A.S. and Yaglom, A.M., {\it Statistical Fluid Mechanics}, 1975, Vol. 2
(MIT Press: Cambridge, Massachusetts).

    Orszag, S.A., Analytical theories of turbulence. {\it J. Fluid
Mech.}, 1970, {\bf 41}, 363-386.

    Ossendrijver, M., The solar dynamo. {\it Astron. Astrophys. Rev.},
2003, {\bf 11}, 287-367.

    Parker, E., {\it Cosmical Magnetic Fields}, 1979 (Oxford University
Press: New York).

    Pipin, V.V., The mean electro-motive force, current- and
cross-helicity under the influence of rotation, magnetic field and
shear. {\it ArXiv: astro-ph/0606265}, 2006, v3.

    Pouquet, A., Frisch, U. and Leorat, J.,  Strong MHD turbulence and
the nonlinear dynamo effect. {\it J. Fluid Mech.}, 1976, {\bf 77},
321-354.

    R\"{a}dler, K.-H., Kleeorin, N. and Rogachevskii, I., The mean
electromotive force for MHD turbulence: the case of a weak mean
magnetic field and slow rotation. {\it Geophys. Astrophys. Fluid
Dynam.}, 2003, {\bf 97}, 249-274.

    R\"{a}dler K.-H., Stepanov R., Mean electromotive force due to
turbulence of a conducting fluid in the presence of mean flow. {\it
Phys. Rev. E}, 2006,  {\bf 73}, 056311 (1-15).

    Roberts, P.H. and Soward, A.M., Dynamo theory. {\it Annu. Rev.
Fluid Mech.}, 1992, {\bf 24}, 459-512.

    Rogachevskii, I. and  Kleeorin, N., Electromotive force for an
anisotropic turbulence: intermediate nonlinearity {\it Phys. Rev.
E}, 2000, {\bf 61}, 5202-5210.

    Rogachevskii, I. and  Kleeorin, N., Electromotive force and
large-scale magnetic dynamo in a turbulent flow with a mean shear.
{\it Phys. Rev. E}, 2003, {\bf 68}, 036301 (1-12).

    Rogachevskii, I. and  Kleeorin, N., Nonlinear theory of a
"shear-current" effect and mean-field magnetic dynamos. {\it Phys.
Rev. E}, 2004, {\bf 70}, 046310 (1-15).

    Rogachevskii, I. and  Kleeorin, N., Effects of differential
and uniform rotation on nonlinear electromotive force in a turbulent
flow. {\it ArXiv: astro-ph/0407375}, 2006, v3.

    Rogachevskii, I.,  Kleeorin, N., Chernin A. D. and Liverts
E., New mechanism of generation of large-scale magnetic fields in
merging protogalactic and protostellar clouds. {\it Astron. Nachr.},
2006, {\bf 327}, 591-594.

    Ruderman, M.S. and Ruzmaikin, A.A., Magnetic field generation
in an anisotropically conducting fluid. {\it Geophys. Astrophys.
Fluid Dynam.}, 1984, {\bf 28}, 77-88.

    R\"{u}diger, G. and Kichatinov, L.L., Do mean-field dynamos in
nonrotating turbulent shear-flows exist? {\it Astron. Nachr.}, 2006,
{\bf 327}, 298-303.

    Ruzmaikin, A.A., Shukurov, A.M. and Sokoloff, D.D., {\it Magnetic
Fields of Galaxies}, 1988 (Kluwer Academic: Dordrecht).

    Stix, M., {\it The Sun: An Introduction}, 1989 (Springer:
Berlin and Heidelberg).

    Urpin, V., Turbulent dynamo action in a shear flow, {\it Monthly Not.
Roy. Astron. Soc.}, 1999a, {\bf 308}, 741-744.

    Urpin, V., Mean electromotive force and dynamo action in a turbulent
flow. {\it Astron. Astrophys.}, 1999b, {\bf 347}, L47-L50.

    Vishniac, E.T. and Cho, J., Magnetic helicity conservation and astrophysical
dynamos. {\it Astrophys. J.}, 2001, {\bf 550}, 752-760.

    Zeldovich, Ya.B.,  Ruzmaikin, A.A. and Sokoloff, D.D., {\it
Magnetic Fields in Astrophysics}, 1983 (Gordon and Breach: New
York).

    Zhang, H., Sokoloff, D., Rogachevskii, I.,  Moss, D., Lamburt, V.,
Kuzanyan, K. and Kleeorin, N., The Radial distribution of magnetic
helicity in solar convective zone: observations and dynamo theory.
{\it Monthly Not. Roy. Astron. Soc.}, 2006,  {\bf 365}, 276-286.

}

\bigskip

\appendix

\section{Threshold for generation of mean vorticity in a sheared turbulence}

\renewcommand{\theequation}
           {A.\arabic{equation}}

The problem of generation of the mean magnetic field in a turbulence
with large-scale velocity shear is similar to that of generation of
mean vorticity in a sheared turbulence. Indeed, let us discuss the
generation of the mean vorticity in a turbulence with a large-scale
linear velocity shear, ${\bf U} = (0, Sx, 0)$  and $ {\bf W} =
(0,0,S)$. Perturbations of the mean vorticity $\tilde{{\bf W}} =
(\tilde{W}_x(z), \tilde{W}_y(z), 0)$ are determined by the following
equations:
\begin{eqnarray}
{\partial \tilde{W}_x \over \partial t} &=& S \tilde{W}_y +
\nu_{_{T}} \tilde{W}''_x  \;,
\label{VE2}\\
{\partial \tilde{W}_y \over \partial t} &=& - \sigma_{_{W}} \, S \,
l_0^2 \, \tilde{W}''_x + \nu_{_{T}} \tilde{W}''_y  \;, \label{VE3}
\end{eqnarray}
(see for details Elperin {\it et al.} 2003), where $\nu_{_{T}}$ is
the turbulent viscosity, $ \tilde{W}''_i = \partial^2 \tilde{W}_i /
\partial z^2$ and the parameter $\sigma_{_{W}}$ is given by
Eq.~(\ref{VE1}) below. The first term, $S \tilde{W}_y$, in the right
hand side of Eq.~(\ref{VE2}) determines a skew-induced generation of
perturbations of the mean vorticity $\tilde{W}_x$ by deflection of
the equilibrium mean vorticity ${\bf W}$, where $\tilde{\bf U}$ are
the perturbations of the mean velocity. In particular, the mean
vorticity $ \tilde{W}_x {\bf e}_x $ is generated from $ \tilde{W}_y
{\bf e}_y $ by equilibrium shear motions with the mean vorticity
${\bf W}= S \, {\bf e}_z ,$ i.e., $S\, \tilde{W}_y {\bf e}_x \propto
({\bf W} {\bf \cdot} \bec{\nabla}) \tilde{U}_x {\bf e}_x \propto
\tilde{W}_y {\bf e}_y \times {\bf W}$. Here ${\bf e}_x$, ${\bf e}_y$
and ${\bf e}_z$ are the unit vectors along $x$, $y$ and $z$ axis,
respectively. On the other hand, the first term, $- \sigma_{_{W}} \,
S \, l_0^2 \, \tilde{W}''_x$, in the right hand side of
Eq.~(\ref{VE3}) determines a Reynolds-stress-induced generation of
perturbations of the mean vorticity $\tilde{W}_y$ by turbulent
Reynolds stresses. This implies that the mean vorticity $\tilde{W}_y
{\bf e}_y$ is generated by an effective anisotropic viscous term $
\propto - l_0^2 \, \Delta \, (\tilde{W}_x {\bf e}_x {\bf \cdot}
\bec{\nabla}) \, U(x) {\bf e}_y \propto - l_0^2 \, S \,
\tilde{W}''_x {\bf e}_y .$ This mechanism of the generation of
perturbations of the mean vorticity $\tilde{W}_y {\bf e}_y $ can be
interpreted as a stretching of the perturbations of the mean
vorticity $\tilde{W}_x {\bf e}_x$ by the equilibrium shear motions
${\bf U}= S \, x \, {\bf e}_y$ during the turnover time of turbulent
eddies (see Elperin {\it et al.} 2003).

Note that equations~(\ref{VE2})-(\ref{VE3}) for the perturbations of
the mean vorticity are very similar to
equations~(\ref{E2})-(\ref{E3}) for the perturbations of the mean
magnetic field in a sheared turbulence. The growth rate $\gamma$ of
the instability of the perturbations of the mean vorticity is given
by
\begin{eqnarray}
\gamma = S \, l_0 \, K_z \, \sqrt{\sigma_{_{W}}} - \nu_{_{T}} \,
K_z^2 \; . \label{VE5}
\end{eqnarray}
The form of the growth rate~(\ref{VE5}) of the perturbations of the
mean vorticity is also very similar to the growth rate~(\ref{IM10})
of the mean magnetic field due to the shear-current effect. On the
other hand, the magnetic dynamo instability is different from the
instability of the perturbations of the mean vorticity although they
are governed by similar equations. The mean vorticity $ \tilde{\bf
W} = \bec{\nabla} {\bf \times} \tilde{\bf U} $ is directly
determined by the velocity field $ \tilde{\bf U}$, while the
magnetic field depends on the velocity field through the induction
equation and Navier-Stokes equation.

In the present study we derived a more general form of the parameter
$\sigma_{_{W}}$ for an arbitrary scaling of the correlation time
$\tau(k)$ of the turbulent velocity field. The parameter
$\sigma_{_{W}}$ is derived using Eq.~(21) of the paper by Elperin
{\it et al.} (2003). It is given by
\begin{eqnarray}
\sigma_{_{W}} = {4 I_0 \over 45} \, \biggl[2 \, {I^2 \over I_0^2} +
43 \, {I \over I_0} + 63 \biggr] \;, \label{VE1}
\end{eqnarray}
where $I$ and $I_0$ are determined by Eq.~(\ref{VE4}). The
instability depends on the correlation time $\tau(k)$. In
particular, when $\tau(k) \propto k^{-\mu}$, the ratio $I/I_0=-\mu$,
and the criterion of the instability reads $2 \, \mu^2 - 43 \, \mu +
63 > 0 $, i.e., the instability is excited when $0 \leq \mu < 1.58$
and $\mu > 19.9$. Note that the condition $\mu > 19.9$ is not
realistic.

When $\tau(k) \propto k^{1-q}$, we recover the result obtained by
Elperin {\it et al.} (2003), i.e., $\sigma_{_{W}} = 4 (2 q^2 - 47 q
+ 108) / 315$. In particular, for the Kolmogorov scaling, $\tau(k)
\propto k^{-2/3}$ (i.e., for $q=5/3$), we arrive at $\sigma_{_{W}}
\approx 0.45$.

For small hydrodynamic Reynolds numbers, the scaling $\tau(k) \sim
1/(\nu k^2)$, the ratio $I/I_0=-2$ (i.e., $\mu = 2$), and the
parameter $\sigma_{_{W}}=-4/9 < 0$. This implies that the
instability of the perturbations of the mean vorticity does not
occur in agreement with the recent results by R\"{u}diger and
Kichatinov (2006). They have not found the instability of the
perturbations of the mean vorticity in a random flow with
large-scale velocity shear using the second order correlation
approximation (SOCA). Note that this approximation is valid only for
small hydrodynamic Reynolds numbers [see discussion in section 3
after Eq.~(\ref{M10})].

\end{document}